\documentclass[aps,prl,superscriptaddress,amsmath,amssymb,aps,twocolumn,]{revtex4-2}
\usepackage{graphicx}
\usepackage{dcolumn}
\usepackage{bm}
\usepackage{latexsym,epsfig}
\usepackage{graphicx}
\usepackage{verbatim}
\usepackage{comment}
\usepackage{amsmath}
\usepackage{amssymb}
\usepackage{stmaryrd}
\usepackage{color}
\usepackage{epstopdf}
\usepackage{grffile}
\usepackage[normalem]{ulem}
\usepackage{float}
\usepackage{tikz}
\usetikzlibrary{quantikz}
\DeclareGraphicsExtensions{.eps}

\newcommand{\beq}{\begin{equation}}
\newcommand{\eeq}{\end{equation}}
\newcommand{\bea}{\begin{eqnarray}}
\newcommand{\eea}{\end{eqnarray}}
\newcommand{\ben}{\begin{eqnarray*}}
\newcommand{\een}{\end{eqnarray*}}
\newcommand{\bfig}{\begin{figure}}
\newcommand{\efig}{\end{figure}}

\usepackage{hyperref}
\hypersetup{
    colorlinks=true,      
    urlcolor=blue,
    citecolor=blue,
    linkcolor=blue
}

\begin{document}
\title{Floquet-induced suppression of thermalization in a quasiperiodic Ising chain}

\author{Biswajit Paul}
\thanks{These authors contributed equally.}
\affiliation{School of Physical Sciences, National Institute of Science Education and Research, Jatni 752050, India}
\affiliation{Homi Bhabha National Institute, Training School Complex, Anushaktinagar, Mumbai 400094, India}

\author{Nilanjan Roy}
\thanks{These authors contributed equally.}
\affiliation{School of Physical Sciences, National Institute of Science Education and Research, Jatni 752050, India}
\affiliation{Homi Bhabha National Institute, Training School Complex, Anushaktinagar, Mumbai 400094, India}

\author{Tapan Mishra}
\email{mishratapan@gmail.com}
\affiliation{School of Physical Sciences, National Institute of Science Education and Research, Jatni 752050, India}
\affiliation{Homi Bhabha National Institute, Training School Complex, Anushaktinagar, Mumbai 400094, India}

\date{\today}

\begin{abstract}
Many-body localized (MBL) systems are known to thermalize in periodically driven systems. In this work, we demonstrate that under proper driving protocol, this thermalization this thermalization can be resisted such that the MBL phase turns into a non-ergodic extended phase, known as the many-body critical (MBC) phase.  
Considering a kicked quasiperiodic Ising chain, we show that while at high-frequency driving the ergodic, MBL, and the MBC phases coexist, at moderate driving frequencies the MBL phase is completely suppressed and the MBC phase proliferates in the parameter space. Using quasienergy statistics, Floquet eigenstates, autocorrelation dynamics, and entanglement growth, we characterize the emergent phases and identify non-monotonic signatures revealing richness of the nonergodic phases. Our results establish Floquet driving as a powerful route to stabilizing nonergodic extended many-body phases beyond the conventional Floquet-MBL paradigm.
\end{abstract}

\maketitle
{\em Introduction.--} 
Periodically driven quantum systems offer a versatile means to engineer nonequilibrium quantum phases that extend beyond their static counterparts ~\cite{Oka2009, Lindner2011, Cayssol2013, Bukov2015}. A generic many-body quantum system typically thermalizes, absorbing energy from the external driving to eventually reach a featureless infinite-temperature or fully mixed state. One mechanism to resist this Floquet heating is integrability, where the system possesses an extensive number of local integrals of motion that limit the energy absorption to only a finite amount. However, such systems are notably susceptible to integrability-breaking perturbations, such as disorder and interactions.

The alternative escape route is many-body localization (MBL), observed in strongly disordered many-body systems, which is argued to avoid thermalization~\cite{abanin2019colloquium,abanin2021distinguishing,sierant2025many,Iyer2013, Schreiber2015, Khemani2017QP,mace2019multifractal}. Intriguingly, under periodic driving, MBL is protected only in the high-frequency limit, where the system is effectively described by a time-averaged Floquet Hamiltonian, and Floquet heating is exponentially suppressed. Consequently, a disorder-induced ergodic-MBL transition is possible in this regime, which has been demonstrated both theoretically ~\cite{Abanin2016, Lazarides2015, Ponte2015} and experimentally ~\cite{bordia2017periodically}. However, reducing the driving frequency induces resonant processes and drive-induced hybridization, leading the MBL system to transition into a thermal phase~\cite{DAlessio2014, Ponte2015PRL, Abanin2017,Lazarides2015}.

In this work, we demonstrate that a periodically driven one-dimensional quasiperiodic lattice, under specific conditions, does not fully thermalize from the MBL phase. Instead, a robust and extensive non-ergodic regime, termed the Many-Body Critical (MBC) regime, emerges. Employing an Ising chain subject to quasiperiodically modulated longitudinal and transverse fields, we reveal that at high driving frequencies, the system's phase diagram, as anticipated, mirrors that of a static (time-averaged) Hamiltonian. This diagram features ergodic, MBL, and MBC regimes, depending upon the quasiperiodic modulation strengths. Significantly, in the low-frequency driving limit, the MBL phase vanishes entirely, leading to a broad MBC phase distinctly separated from the ergodic phase by a sharp boundary. Notably, within the low-frequency driving regime, a dynamical crossover line is identified within the MBC phase. Along this line, the system exhibits maximal suppression of delocalization as the longitudinal field strength increases, a feature absent in both the static and the high-frequency regimes of our driven model. Our findings present a counter-intuitive observation within the finite system sizes accessible through exact numerics, contrasting with the conventional understanding of the MBL-ergodic transition under slow driving, and suggest the possibility of an MBL-to-MBC crossover upon decreasing the driving frequency.

\begin{figure}
\begin{center}
\includegraphics[width=1.0\columnwidth]{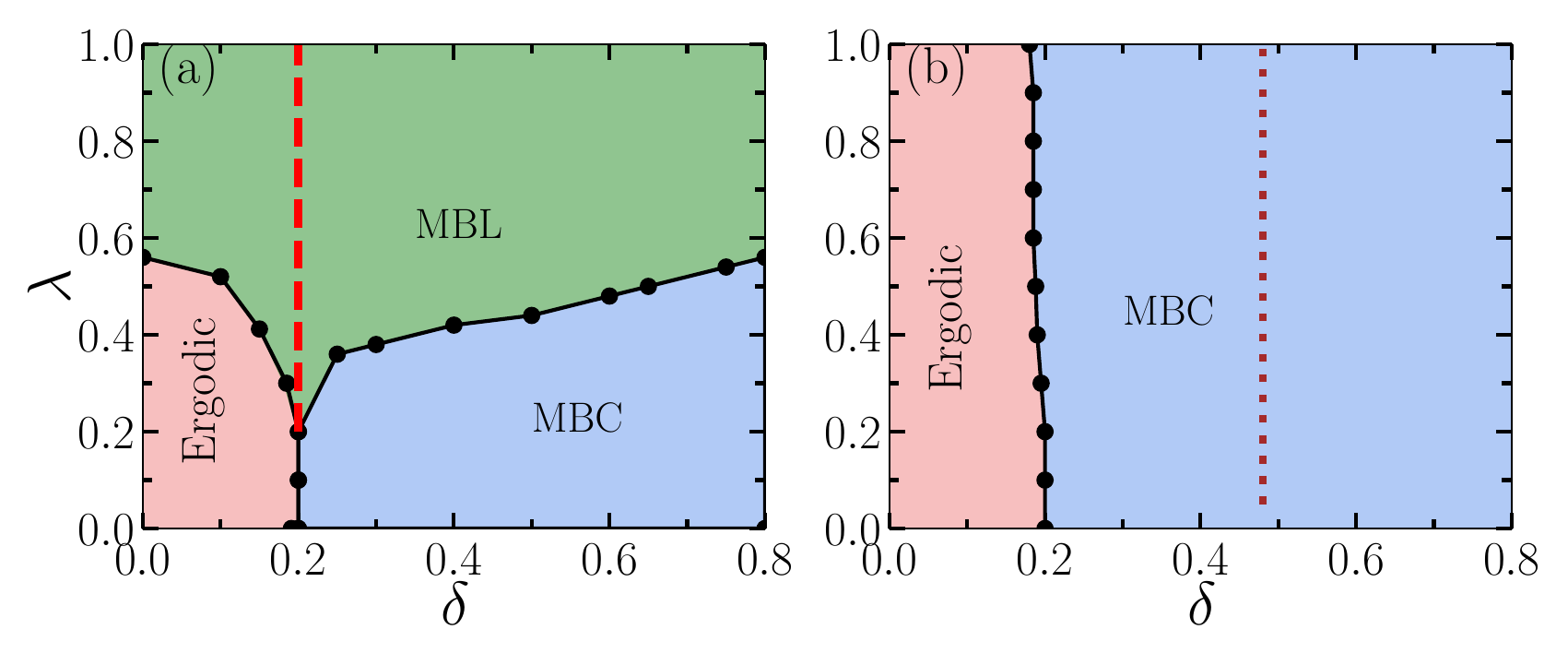}
\end{center}
\caption{Phase diagram: (a) At driving period $T=0.02\pi$ (high frequency), the ergodic, MBC, and MBL phases are separated by solid lines and markers. A Widom-like line emerges from the triple point inside the MBL phase as denoted by the dashed line. (b) At $T=1.5\pi$ (moderate driving frequency), we observe mainly two phases i.e. ergodic and MBC phases which are separated by the solid black line. Deep within the MBC phase, we observe a dynamical crossover line, indicated by a dotted line in the figure.}
\label{fig:phase_diag}
\end{figure}

{\em Model.--}
We study a periodically driven spin-$\frac{1}{2}$ Ising chain, where the dynamics is governed by the time-dependent Hamiltonian;
\begin{equation}
    \hat{H}(t) = 
    \begin{cases}
    \hat{H}_x = \sum_j g_j ~\hat{\sigma}_j^x
        , & \text{if $0\le t< \frac{T}{2}$}\\
        \hat{H}_z = J\sum_j \hat{\sigma}_i^z\hat{\sigma}_{j+1}^z+\sum_i h_j\hat{\sigma}_j^z
        , & \text{if $\frac{T}{2}\le t<T $}.
    \end{cases}
    \label{eq:drive}
\end{equation}
Here, $h_j = \lambda\cos(2\pi b j+\phi)$ is the strength of quasi-periodic longitudinal field, and $g_j = \mu+\delta\cos(2\pi b j+\phi)$ is the strength of transverse field at site $j$, $b = \frac{\sqrt{5}-1}{2}$ is the inverse golden ratio and $\phi\in [0, 2\pi)$ are the arbitrary phase. $\hat{\sigma}^{a}$, $a\in{x, y, z}$ are the Pauli matrices. For this system, the unitary evolution operator for a single Floquet cycle can be defined as
\begin{align}
    \hat{U}_F &= e^{-\frac{iJT}{2}\sum_j\hat{\sigma}_j^z\hat{\sigma}_{j+1}^z} e^{-\frac{iT}{2}\sum_jh_j\hat{\sigma}_j^z}e^
    {-\frac{iT}{2}\sum_jg_j\sigma_j^x}\nonumber\\
    &=\prod_j^{L-1} R^j_{ZZ}(JT)\prod_j^L R_Z^j(h_jT)\prod_j^L R_X^j(g_jT).
    \label{eq:unitary_opr}
\end{align}
where $\hat{U}_F$ is written as a product of unitary operators $R_{ZZ}$, $R_X$ and $R_Z$ where $R_X$ and $R_Z$ obey the rotational periodicity of $\pi$. Hence, for all the calculations, we consider periodic boundary condition (PBC) and  choose the values of $J$, $\mu$ and ranges of $T$, $\lambda$ and $\delta$ such that $\frac{h_jT}{2}<\pi$ and $\frac{g_jT}{2}<\pi$. 
Low dimensional quantum systems with quasiperiodic coupling and disorder have been well studied in the context of the static model, where the existence of MBC phase has been reported that emerges from interaction-induced mixing of critical states in the non-interacting limit~\cite{wang2021many,roy2025many}. 
In the following, we reveal the onset of the MBC phase and its stability in a periodically driven system.

{\em Phase diagram.--}
We begin with a brief discussion on the phase diagram shown in Fig.~\ref{fig:phase_diag}(a-b) in the ranges $0<\delta<0.8$  and $0<\lambda<1$ for high $(\omega=100)$ and moderately low $(\omega=4/3)$ frequencies, respectively, for fixed $J=\mu=0.2$ .
For the high frequency drive, we obtain three phases, namely; the ergodic, MBL and MBC phases as shown in  Fig.~\ref{fig:phase_diag}(a)). The ergodic phase appears when the disorder strength $\lambda$ is weak and $\delta<\mu$. However, the MBC phase appears for the weak disorder limit and when $\delta \geq \mu$. With increase in $\lambda$, both the phases transition to an MBL phase resulting in two phase boundaries, i.e., the ergodic-MBL (black circles) and MBC-MBL (black squares). Note that a similar phase diagram can be obtained from static Hamiltonian, $\hat{H}_F = (\hat{H}_z + \hat{H}_x)/2$, derived from high-frequency Magnus expansion of the Floquet Hamiltonian $\hat{H}_F = \frac{i}{T}\ln(\hat{U}_F)$~\cite{suppl}.
Interestingly, however, for driving at a moderately low frequency both the ergodic and MBC phases are enhanced by completely suppressing the MBL region. As a result, the ergodic-MBL and MBC-MBL phase boundaries merge to become a sharp phase boundary between the ergodic and MBC phase. It is important to note that while the periodic driving turns the MBL phase ergodic as expected (for $\delta<0.2$), counterintuitively the MBC phase remains robust in the large $\delta>0.2$ and this is our central result. In the following we will characterise these phases by exact numerical calculations. 

\begin{figure}[t!]
\begin{center}
\includegraphics[width=1.0\columnwidth]{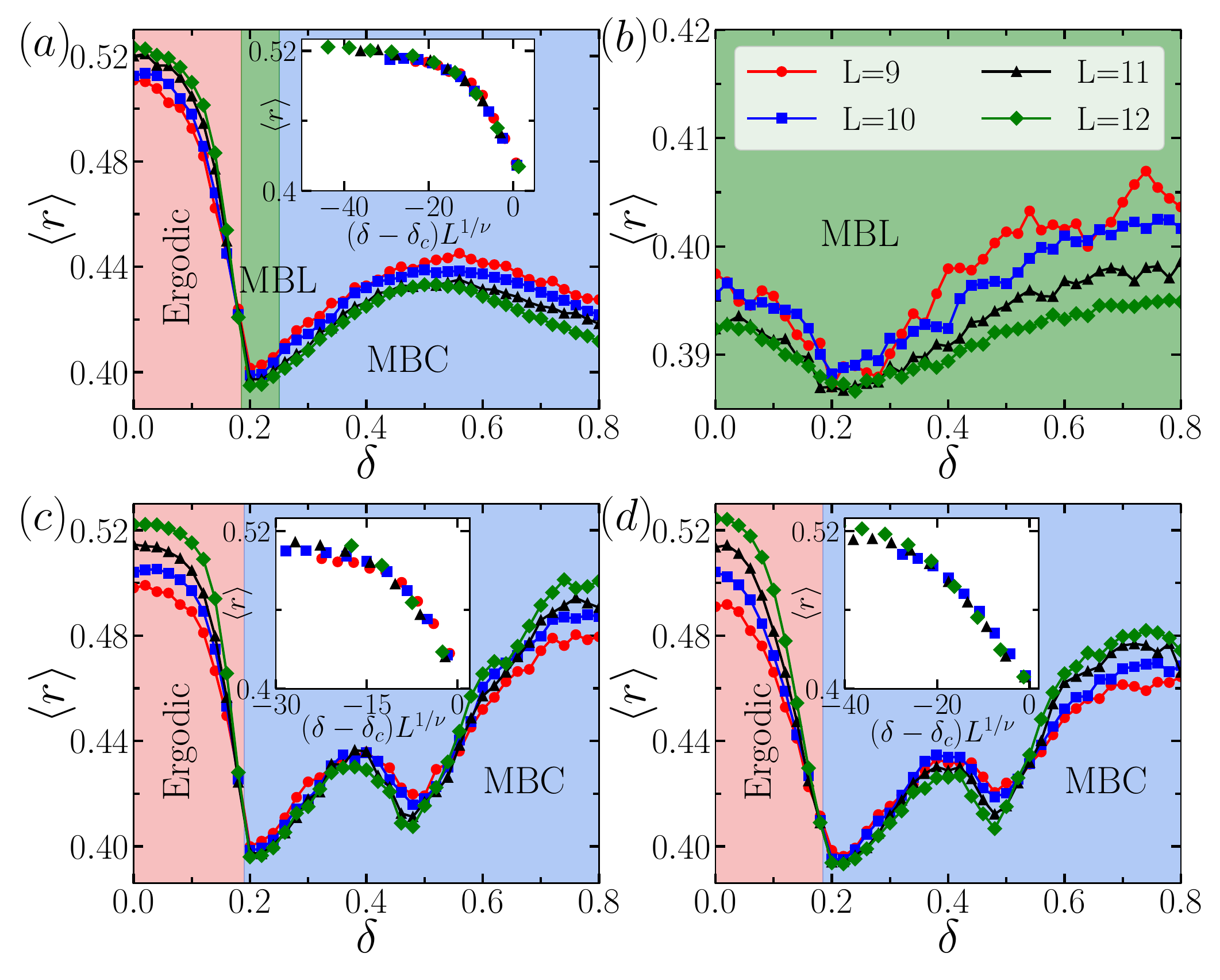}
\end{center}
\caption{In figures (a) and (b) the averaged-LSR ($\langle r\rangle$) is plotted as a function of $\delta$ for increasing system sizes $L$, for $\lambda=0.3$ and $\lambda=0.8$, respectively, for fixed $T=0.02\pi$. Figures (c) and (d) show similar plots, but now for $T=1.5\pi$ for increasing system sizes while keeping the other parameter values same as in Fig.~(a) and (b). Inset shows scaling and collapse of data as $\langle r \rangle \sim (\delta -\delta_c)L^{1/\nu}$.}
\label{fig:LSR_IPR}
\end{figure}

{\em Numerical Results.-}
We first probe the phases by analyzing the quasi-energy level statistics of the Floquet Hamiltonian $\hat{H}_F$  through the level spacing ratio~(LSR). The LSR of the quasi-energy spectrum, $\epsilon_j T \in [0, 2\pi)$  is defined as
$r = \frac{\min(\Delta_j, \Delta_{j+1})}{\max(\Delta_j, \Delta_{j+1})}$,
where $\Delta_j = \epsilon_{j+1} - \epsilon_j$ denotes the spacing between consecutive quasi-energy levels arranged in ascending order. In the ergodic phase, the level statistics follow the Wigner-Dyson distribution, namely, the circular orthogonal ensemble (COE)~\cite{mehta2004random,DAlessio2014,regnault2016floquet}, while in the MBL phase they obey Poisson statistics~\cite{Pal2010,khemani2016phase}. Accordingly, the averaged LSR, denoted by $\langle r \rangle$ approaches $\langle r \rangle \approx 0.526$ in the ergodic regime and $\langle r \rangle \approx 0.386$ in the localized regime. Here $\langle ..\rangle$ represents the average of LSR over the entire spectrum with random phase offsets $\phi=400$, $400$, $200$, and $100$ random phase offsets $\phi$ for system sizes  $L=9$, $10$, $11$, and $12$, respectively. In the critical regime, $\langle r \rangle$ assumes intermediate values between these two limits. 
In  Fig.~\ref{fig:LSR_IPR}(a) and Fig.~\ref{fig:LSR_IPR}(b), we plot $\langle r \rangle$ as a function of $\delta$ for a fixed (high-frequency) driving period $T=0.02\pi$ ($\omega = 100$) and for two values of $\lambda=0.3$ and $\lambda=0.8$, respectively. We observe that for $\delta \lesssim \mu=0.2$ the system is in the ergodic phase which can be attributed  to the spin flip term through which the system explores all configurations in the Hilbert-space with equal probability. With increase in $\delta$, the system enters into a small window $(0.2<\delta<0.25)$ of the MBL phase where $\langle r\rangle\sim 0.386$.
For $\delta>0.25$, $0.386<\langle r\rangle<0.526$ indicating the presence of an MBC phase. This results in an ergodic-MBL-MBC transition as function of $\delta$ (compare Fig.~\ref{fig:phase_diag}(a). In the limit of very small $\lambda<0.2$, the MBL phase does not appear, leading to an ergodic-MBC transition at $\delta=\mu=0.2$ (not shown). This happens due to the sudden appearances of near-zero $g_j$ terms, as discussed earlier, which maximizes at $\delta=\mu$ and persists for $\delta>\mu$ making the possibility of ergodic states unfeasible. 
At higher values of $\lambda$ ($\lambda=0.8$), the system favours to be in MBL phase for all the values of $\delta$. This can be seen from the behaviour of $\langle r\rangle$ as a function of $\delta$ in Fig.~\ref{fig:LSR_IPR}(b) which shifts toward Poisson statistics with increasing system size. Interestingly, we find a sharp dip in $\langle r\rangle$ at $\delta=\mu$ tracing the presence of the Widom-like line along which the eigenstates are maximally localized for fixed $\lambda$, as also depicted in Fig.~\ref{fig:phase_diag}(a) (red dashed line). This maximal localization increases as one approaches the triple point from which the Widom line emerges.

The scenario becomes particularly intriguing at the moderate driving frequency regime. In this regime, one needs to consider higher order terms in the Magnus expansion  representing the quasiperiodic long-range couplings in the Floquet Hamiltonian, beyond the short range static Hamiltonian in the high-frequency regime~\cite{suppl}. In Fig.~\ref{fig:LSR_IPR}(c) and Fig.~\ref{fig:LSR_IPR}(d), we show the LSR as a function of $\delta$ for the fixed (low-frequency) driving period $T=1.5\pi$ ($\omega = 4/3$) and two values of $\lambda=0.3$ and $\lambda=0.8$, respectively.
Both the figures show a transition from the ergodic to MBC phase at $\delta\approx\mu=0.2$, showing almost no dependence and very small dependence on the lower and higher values of $\lambda$, respectively (also compare  Fig.~\ref{fig:phase_diag}(b)). 
Deep in the MBC phase, we obtain a non-monotonic behavior (a local minimum) of $\langle r \rangle$ around $\delta \sim 0.48$, which is consistent with the location of the crossover line $T(\mu + \delta)\approx \pi$, denoted as the dotted vertical line  in Fig.~\ref{fig:phase_diag}(b). Along this line the system shows localization tendency within the MBC phase which signifies an effectively enhanced non-uniformity in mixing due to spin-flipping term.
Moreover, from both the figures in low-frequency regime, we observe a rise and a crossover, in terms of system-size dependence in $\langle r\rangle$, at around $\delta > 0.55$, indicating a tendency of delocalization within the critical regime. However, the rate of increase in the values of $\langle r\rangle$ in this regime decreases with increasing $\lambda$.

The insets of Fig.~\ref{fig:LSR_IPR}(a) and Fig.~\ref{fig:LSR_IPR}(c-d) show the collapse of data obtained in the ergodic side of the ergodic-MBL and ergodic-MBC transitions, respectively, following a linear scaling ansatz $\langle r\rangle\sim f(L/\xi)$, where $\xi$ is the correlation length. The collapse of data shows that $\langle r\rangle\sim (\delta - \delta_c) L^{1/\nu}$ revealing the power-law divergence of the correlation length at the transition point as $\xi\sim (\delta-\delta_c)^{-\nu}$ where $\delta_c$ is the critical point. From our analysis, we obtain $(\delta_c=0.185, \nu = 0.37)$, $(\delta_c=0.19, \nu = 0.45)$, $(\delta_c=0.185, \nu = 0.45)$ for Fig.~\ref{fig:LSR_IPR}(a), Fig.~\ref{fig:LSR_IPR}(c) and Fig.~\ref{fig:LSR_IPR}(d), respectively.
It is important to note that the MBL phase of the high-frequency regime becomes an MBC phase in the limit of low-frequency driving, indicating proliferation of the MBC phase with decreasing driving frequency. 
Interestingly, lowering the driving frequency, even further, shows the   shifting of the crossover line towards a lower value of $\delta$ tracing $(\mu+\delta)T\approx \pi$. However, the ergodic-MBC transition still happens at $\delta_c=\mu$, irrespective of $\lambda$. This robust MBC phase appears for $\delta\geq\mu$ due to the appearance of near-zero numbers in the distribution of $g_j$ (i.e. the strength of spin-flipping terms), which maximizes at $\delta=\mu$ and continues to persist for $\delta>\mu$~\cite{suppl}. Such a non-uniform distribution of $g_j$ leads to Floquet eigenstates with highly non-uniform connections in Fock-space, thus, protecting the MBC phase, even at the low-frequency regime. 
\begin{figure}[t]
\begin{center}
\includegraphics[width=1.\columnwidth]{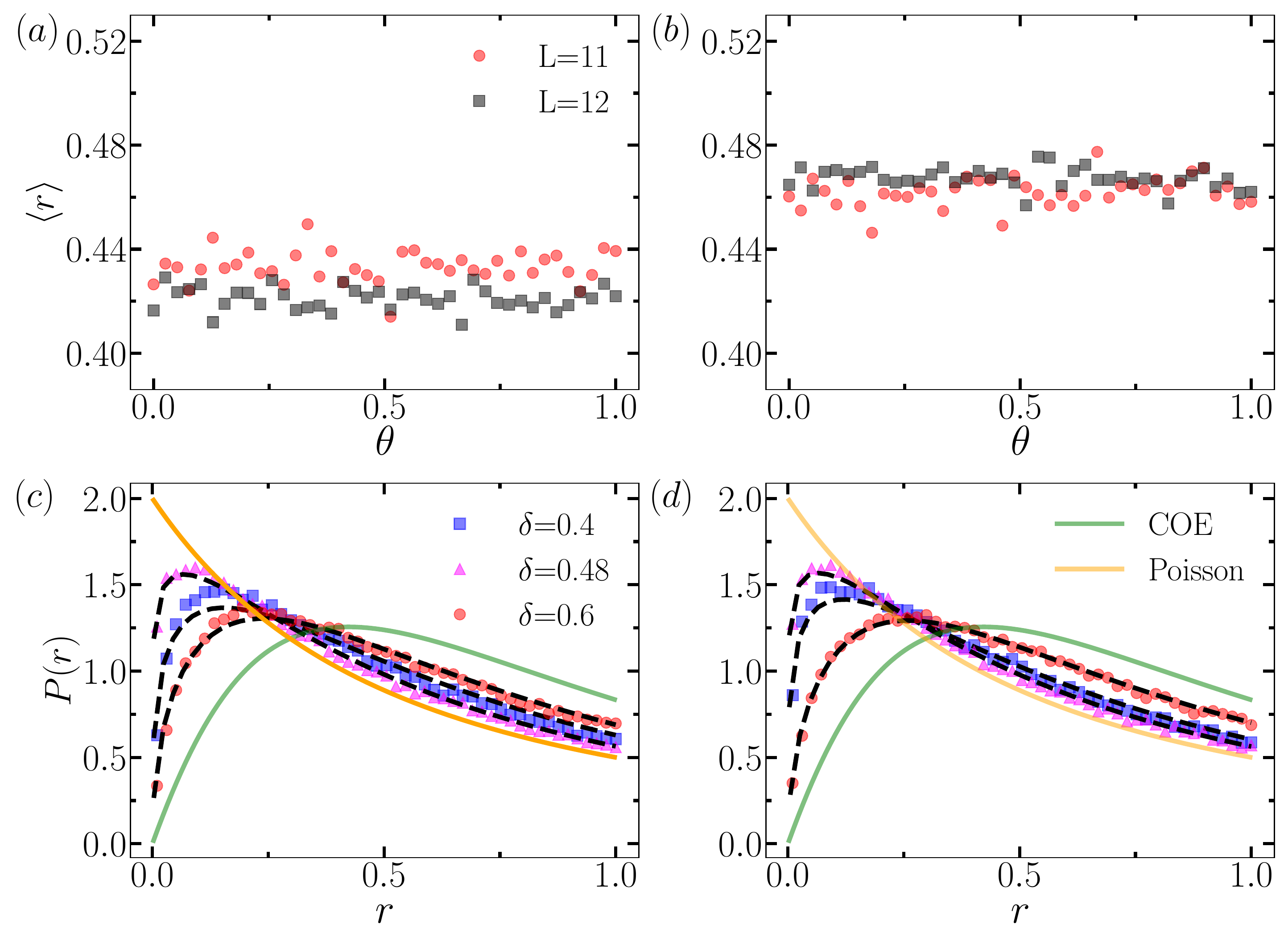}
\end{center}
\caption{Figures (a) and (b) show energy- resolved $\langle r \rangle$ for $\delta=0.4$ and $\delta=0.6$, respectively for system sizes $L=11,12$, while keeping $J=0.2$, $\mu=0.2$, and $\lambda=0.8$ fixed. Figures (c) and (d) show the probability distribution~($P(r))$ for $\lambda=0.4$ and $\lambda=0,6$, respectively, for increasing values of $\delta$. We keep $J=0.2$, $\mu=0.2$, and $L=12$ fixed for these two figures.}
\label{levelstat}
\end{figure}
To confirm that the entire spectrum of the MBC phase is critical, we show the energy-resolved LSR of the quasi-energies in Figs.~\ref{levelstat}(a) and (b) for $\delta=0.4$ and $0.6$, respectively, at fixed $\lambda=0.8$ and $T=1.5\pi$ (moderate driving frequency). The plot in Fig.~\ref{levelstat}(a) for system size $L=11,12$ reveals that $0.41<\langle r\rangle<0.43$. Similarly, $0.46<\langle r\rangle<0.48$ for $L=12$ in Fig.~\ref{levelstat}(b), which implies that all the Floquet eigenstates are of the nonergodic extended (NEE) nature, intermediate between ergodic and MBL phases. 

We next discuss the probability distributions of level spacing ratio $P(r)$ in the MBC phase. In the MBL phase, $P_\text{P}(r) = \frac{2}{(1+r)^2}$ which corresponds to Poisson value $\langle r\rangle \approx 0.386$ whereas in the ergodic phase $P_\text{COE}(r) = \frac{2}{3}\left(\frac{\sin(\frac{2\pi r}{r+1})}{2\pi r^2}+\frac{1}{(1+r)^2} + \frac{\sin(\frac{2\pi}{r+1})}{2\pi} - \frac{\cos(\frac{2\pi}{r+1})}{r+1} - \frac{\cos(\frac{2\pi r}{r+1})}{r(r+1)} \right) $ which corresponds to COE value $\langle r\rangle \approx 0.526$~\cite{DAlessio2014}. Here for MBC phase we propose $P_\text{c} (r) = \frac{1}{Z_\beta} [P_\text{P}(r)]^{\beta_r} [P_\text{COE}(r)]^{1-\beta_r}$, where $Z_\beta$ is a normalization constant. $\beta_r=1$ and $0$ correspond to Poisson and COE distributions, respectively, shown as the solid lines in Figs.~\ref{levelstat}(c,~d). For moderate frequency corresponding to $T=1.5\pi$, we extract the exponent $\beta_r$ after fitting $P_\text{c}(r)$ with the data points shown as dashed lines in Fig.~\ref{levelstat}(c) and Fig.~\ref{levelstat}(d) which are plotted for $\lambda=0.4$ and $0.6$, respectively, for different values of $\delta$ for each case. From Fig.~\ref{levelstat}(c), we obtain $\beta_r= 0.73,~0.87$, and $0.53$ for $\delta=0.4,0.48$ and $0.6$, respectively, whereas from Fig.~\ref{levelstat}(d), $\beta_r= 0.78, 0.87$, and $0.49$ for the same values of $\delta$ as in Fig.~\ref{levelstat}(c). $\beta_r$ shows maximum value tracing the dynamical crossover line at $\delta\approx0.48$ within the MBC phase. Thus, our study provides with an working expression of $P(r)$ in the MBC phase, the instances of which are rare in quantum systems.

A complementary study of the inverse participation ratio~(IPR) that captures the localization properties of the Floquet eigenstates is discussed in the supplementary materials[\onlinecite{suppl}]. 
The phase boundaries in the phase diagram shown in Fig.~\ref{fig:phase_diag} are primarily identified by comparing the behavior of LSR and IPR.
After getting the insights from the eigenvalues and eigenstates of Floquet Hamiltonian about the phases, we investigate the out-of-equilibrium dynamics to understand their behaviour in the following.

\begin{figure}[t!]
\begin{center}
\includegraphics[width=1.0\columnwidth]{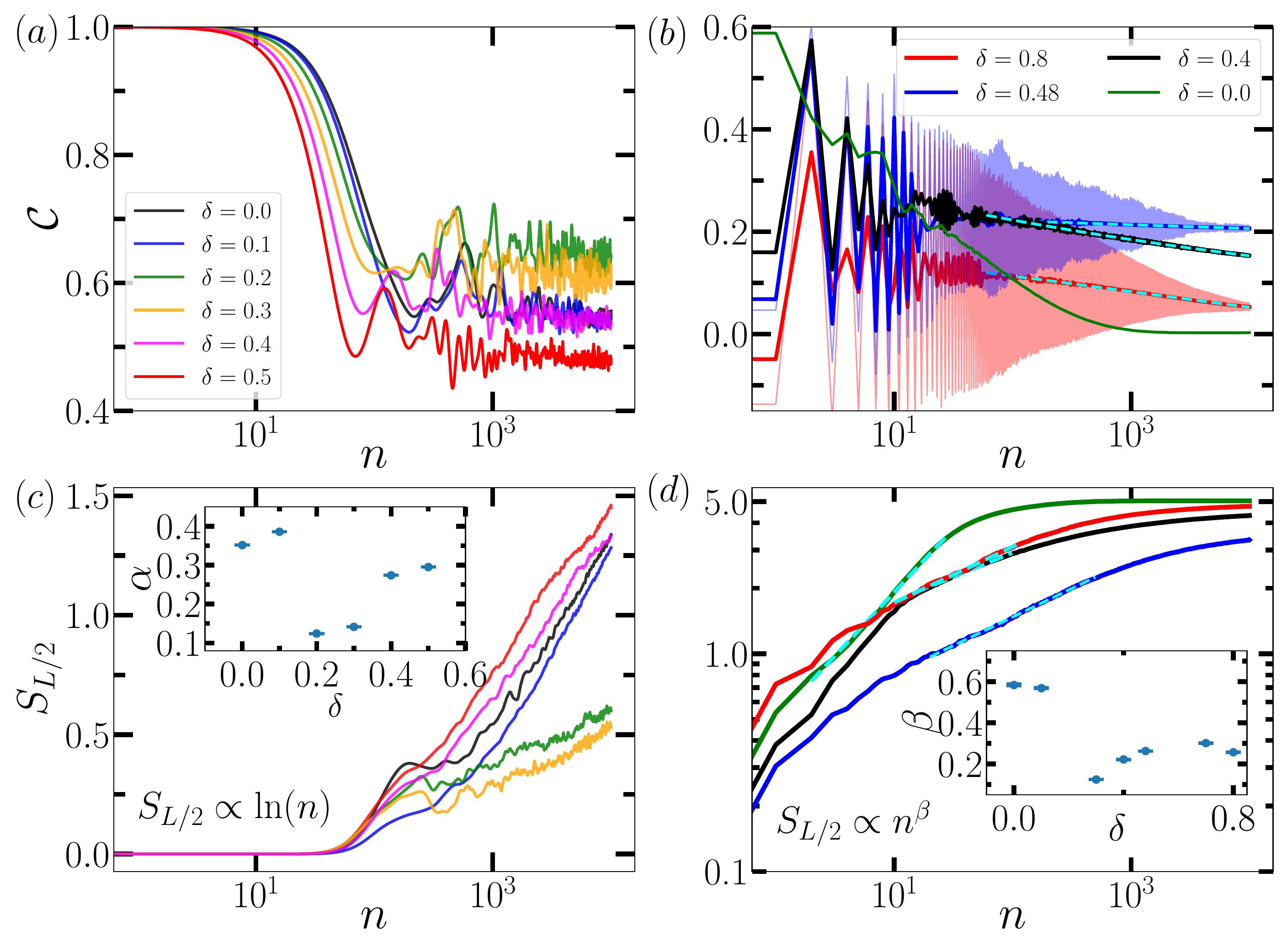}
\end{center}
\caption{Figures (a) and (b) show the autocorrelation $\mathcal{C}$ as a function of the Floquet-step $n$ for driving period $T = 0.02\pi$ and $T = 1.5\pi$, respectively. Figures (c) and (d) illustrate the half-system EE $(S_{L/2})$ as a function of Floquet-step $n$ for the same values of $T$ as in panels (a) and (b), respectively. We keep the other parameter fixed at $\mu = 0.2$, $J = 0.2$ and $\lambda = 0.7$. For all plots $L=16$. The dashed lines in figures (b) and (d) represent the fitted curve. The smooth solid lines for the fluctuating data in (b) are obtained using the spline method (a global interpolation method) implemented with the Python-based SciPy~\cite{DIERCKX1975165, SciPy-NMeth} library. The insets show various exponents as function of $\delta$.}
\label{fig:corr_EE}
\end{figure}
{\em Dynamics.-}
We investigate the dynamics from the behviour of autocorrelation defined as,
$
    \mathcal{C}(nT) = \frac{1}{L}\sum_0^{L-1}\langle \psi(0)|\hat{\sigma}_i^z(nT)\hat{\sigma}_i^z(0)|\psi(0)\rangle
$
We start the dynamics from random Fock states, and for each case we consider a random phase offset $\phi$ to avoid the initial state dependence of the data. 
For an ergodic system, the autocorrelation is expected to vanish rapidly since the system does not retain memory of the initial state~\cite{DAlessio2014}. On the other hand, for the MBL phase, the autocorrelation remains non-zero due to the presence of the local integrals of motion~\cite{DAlessio2014,Abanin2016}. However, in the MBC phase, the picture of local integrals of motion  may be approximately preserved (although not shown explicitly in the literature) and hence auto-correlation is expected to show a slower decay than that in the ergodic phase. In Fig.~\ref{fig:corr_EE}(a) and Fig.~\ref{fig:corr_EE}(b) we show the autocorrelation as a function of Floquet step $n$ for driving period $T=0.02\pi$ and $1.5\pi$, respectively, with fixed $\lambda=0.7$ and for different values of $\delta$. For $T=0.02\pi$ (high frequency), the saturation of $\mathcal{C}(nT)$ in Fig.~\ref{fig:corr_EE}(a) at large values (order of $1$) irrespective of the value of $\delta$, indicates an MBL phase. The saturation value reaches maximum at $\delta=0.2$ tracing the presence of the Widom line. However, the behavior of the system becomes intriguing for the low frequency case (Fig.~\ref{fig:corr_EE}(b)). In this case, when $\delta=0$,  $\mathcal{C}$ decays exponentially and becomes zero which is consistent with the ergodic behavior. However, as we enter the MBC phase (i.e., for $\delta=0.4$), we obtain  that after an initial fluctuations in the first few cycles, the autocorrelation shows a slow power-law decay as $\mathcal{C}\propto n^{-\gamma_p}$. For $\delta=0.4$ the exponent $\gamma_p \approx 0.082$ which reaches minimum ($\gamma_p \approx 0.015$) at around $\delta=0.48$. 
Further increasing $\delta$ beyond the crossover line $(\delta\geq 0.55)$, the decay of $\mathcal{C}$ better fits with a stretched exponential behavior i.e. $\mathcal{C}\propto \exp[-(n/\tau)^{\gamma_e}]$, e.g. at $\delta=0.8$, we find $\gamma_e\approx 0.17$. This behavior is an indication of delocalization tendency of the MBC phase of the system~\cite{suppl}.
We note that beyond $\delta\geq 0.48$ the initial fluctuations of $\mathcal{C}$ (which decrease with $L$) starts persisting for longer cycles which eventually settles down after a large number of cycles. Hence, in order to obtain a better fit we first  smoothen the fluctuating curves in this regime shown as dark solid lines.

To complement the dynamical behaviors from the autocorrelation, we  compute the half-chain entanglement entropy (EE), $S_{L/2}(n T) = -{\rm Tr}\left[\rho_A(nT) \ln (\rho_A(nT) )\right]$, of the bipartite quantum system for the time-evolving quantum states at each stroboscopic interval starting from the random Fock states. Here, $\rho_A(nT) = {\rm Tr}_B(|\psi(nT)\rangle\langle\psi(nT)|)$ is the reduced density matrix for the subsystem $A$ by taking the partial trace over rest of the system denoted by $B$. 
In the ergodic phase, one expects the EE to rapidly grow and saturate near the Page value $\sim (0.5 L \ln{2}-0.5)$~\cite{page_value_1993, page_value_2017}. On the other hand, for the MBL phase it shows the characteristic logarithmic growth~\cite{abanin2019colloquium}.

From Fig.~\ref{fig:corr_EE}(c) we can see a logarithmic growth of $S_{L/2}$ for all values of $\delta$ when $\lambda=0.7$ for high driving frequency, confirming the presence of the MBL phase throughout. We fit the logarithmic growth of $S_{L/2}$ with the function $S_{L/2} = \alpha\ln(n)+c$ and plot $\alpha$ as a function of $\delta$ in the inset of the same figure. The inset shows that $\alpha$ reaches a minimum at $\delta=0.2$ indicating the slowest growth that traces the Widom line in the MBL phase at high-frequency regime. We also study the dynamical behavior of EE at $T=1.5\pi$ (low-frequency), which can be seen from Fig.~\ref{fig:corr_EE}(d). In the log-log plots we fit the linear region of EE growth with the function $S_{L/2}\propto n^\beta$, and plot the values of the power-law exponent in the inset. In the ergodic regime ($\delta<0.2$), the $S_{L/2}$ grows sub-ballistically and within a few Floquet steps it saturates to the Page value, although the growth exponent tends towards the ballistic value with increasing $L$~\cite{bera2019slow}. However, in the MBC phase, we can observe a sub-diffusive behavior $(\beta\leq 0.3)$, i.e., a very slow growth of EE. In this subdiffusive regime, the long time value of EE shows a local minimum near $\delta\approx0.48$, consistent with the presence of the dynamical crossover line. Throughout the MBC phase, after the subdiffusive growth, the EE grows even in a much slower rate towards the saturation which is absent in the ergodic and MBL phases.

{\em Conclusion.-}
In this work, we study a special type of kicked quasiperiodic Ising spin chain to demonstrate absence of thermalization of the MBL phase and an onset of robust and broad non-ergodic critical (MBC) phase which is typically rare in periodically driven systems. 
Moreover, we reveal intricate structures in the phases such as the Widom-like line in high-frequency MBL phase and dynamical crossover line within the low-frequency MBC phase. The nonergodic phases, especially the MBC phase in our model can find useful applications, such as, in controlled quantum information transfer~\cite{zhou2019floquet,geier2021floquet,monroe2025floquet,altland2024qc}, engineering anomalous transport~\cite{extended2020yang} and quantum sensing~\cite{MONTENEGRO20251,mihailescu2025critical}. The phases of our model are directly accessible in existing cold atoms~\cite{bordia2017periodically}, Rydberg arrays~\cite{Bluvstein_2021}, and superconducting qubits\cite{Mi2022, TC2022} platforms. Thus, our work would contribute to both fundamental studies and quantum technology applications.


\bibliography{refs}

\clearpage
\onecolumngrid
\begin{center}
\textbf{Supplementary materials for 
 ``Floquet-induced suppression of thermalization in a quasiperiodic Ising chain"}
\end{center}

\vspace{5mm}

\renewcommand{\appendixname}{}
\renewcommand{\thesection}{{S\arabic{section}}}
\renewcommand{\theequation}{S\arabic{equation}}
\renewcommand{\thefigure}{S\arabic{figure}}

\setcounter{section}{1}
\setcounter{figure}{0}
\setcounter{equation}{0}

\onecolumngrid
\section{\thesection. Numerical method for time evolution}

The dynamics of the spin chain is governed by the unitary Floquet operator given in the main text. The first half of the dynamics is led by the Hamiltonian $\hat{H_z}$ and the other half by $\hat{H}_x$. Note that the Hamiltonian $\hat{H}_z$ and $\hat{H}_x$ are diagonal in $\hat{\sigma}^z$ and $\hat{\sigma}^x$ basis, respectively. We have implemented the fast Hadamard transformation method for time evolution as mentioned in a previous work~\cite{bera2019slow}. The Hadamard transformation transforms the basis from $\sigma^z$ to $\sigma^x$ through the unitary transformation operator $U_H = \bigotimes_{i=0}^{L-1} \frac{1}{\sqrt{2}}\begin{pmatrix}
    1 & 1\\
    1 & -1
\end{pmatrix}$. We implement the Fast Hadamard transformation~\cite{DIERCKX1975165,SciPy-NMeth} to change the basis, which requires $\mathcal{N}\log_2{\mathcal{N}}$ number of operations. The single step of the time evolution operator can be written as,
\begin{equation}
    |\psi((n+1)T)\rangle = U_H U_xU_HU_z|\psi(nT)\rangle.
\end{equation}
Where, $U_z = e^{-i\hat{H}_zT/2}$ and $U_x = U_He^{-i\hat{H}_xT/2}U_H$ are diagonal in $\sigma^z$ and $\sigma^x$, respectively. This method allows efficient and exact time evolution at larger system sizes.

\begin{figure}[b!]
\begin{center}
\includegraphics[width=1\columnwidth]{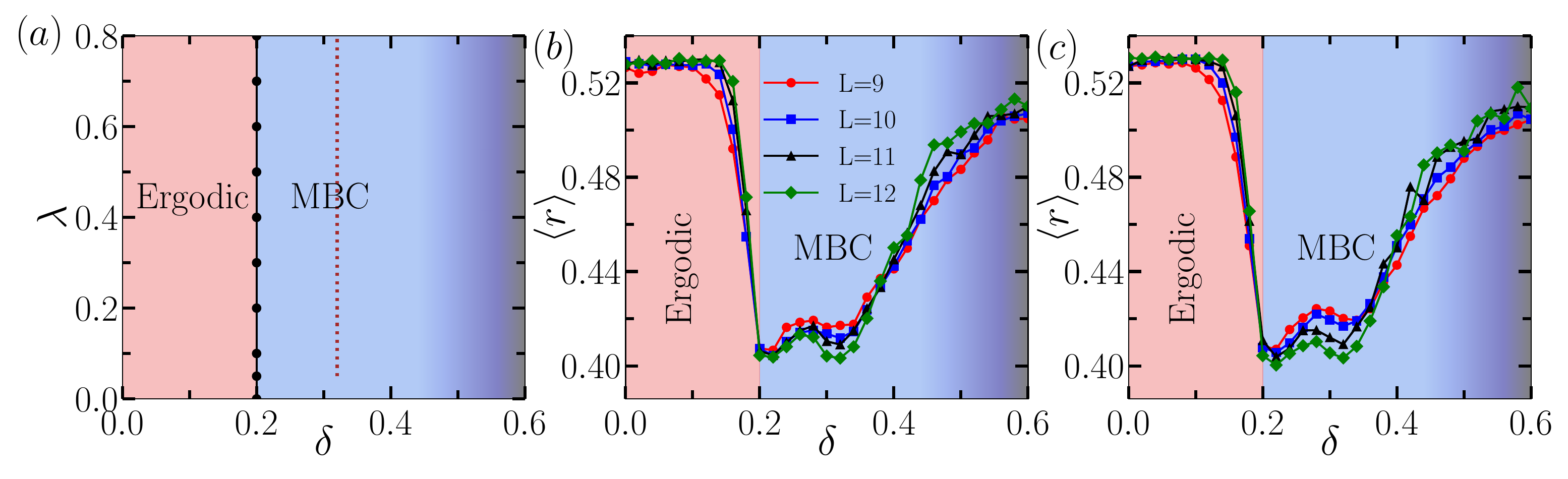}
\end{center}
\caption{ Figure~(a) shows the phase diagram for $T=2\pi$. In figures (b) and (c) averaged-LSR ($\langle r\rangle$) is plotted as a function of $\delta$ for increasing system sizes $L$, for $\lambda=0.3$ and, $\lambda=0.8$, respectively, for fixed $T=2\pi$.}
\label{fig:phase_diag_LSR_2Pi}
\end{figure}

\setcounter{section}{2}
\section{\thesection. Lower-frequency driving analysis}

In addition to the results in the main text in the high and low frequency driving limits, we have studied the system by further lowering the frequency i.e. $\omega=1$ or $T=2\pi$. The phase diagram is plotted in Fig.~\ref{fig:phase_diag_LSR_2Pi}(a), by analyzing the averaged LSR and IPR which will be discussed in the next section. We observe the ergodic phase for $\delta<\mu=0.2$ and the MBC phase for $\delta>0.2$, irrespective of the onsite disorder strength. In the regime of $\delta>0.2$ is critical in nature, we make a blue colour gradient region to indicate the increasing value of LSR. We observe the non-monotonic behavior as a function of $\delta$, with an increasing localization around $\delta\sim 0.32$ within the MBC phase. This non-monotonic behavior essentially traces a crossover line $(\mu + \delta)T\approx \pi$ which is denoted as a dot-dashed line in the phase diagram. Note that the crossover line shift towards lower value of $\delta$ as we reduce the frequency. To closely look into the system we plot the averaged LSR, denoted as $\langle r \rangle$, as a function of $\delta$ for $\lambda = 0.3$ and $0.8$ in Fig.~\ref{fig:phase_diag_LSR_2Pi}(b) and Fig.~\ref{fig:phase_diag_LSR_2Pi}(c), respectively. For the value of $\delta<0.2$, the $\langle r \rangle\simeq 0.53$ indicates, for this parameter range the system in ergodic phase. However, for $0.2<\delta<0.35$ the value of LSR lies near $\langle r \rangle\simeq0.41$ with showing minimum value at around $\delta=0.32$. Further increase in $\delta$ value causes a sharp increase in $\langle r \rangle$ value, which can be seen from both the Fig.~\ref{fig:phase_diag_LSR_2Pi}(b) and Fig.~\ref{fig:phase_diag_LSR_2Pi}(c). When the value of $\delta\sim 0.6$, we can see LSR value nearly saturated at around $\langle r \rangle\simeq 0.51$, which does not scale with the system size as well. 
Hence, the delocalization tendency of Floquet eigenstates within MBC phase in higher-$\delta$ regime increases with lowering the driving frequency.  

\setcounter{section}{3}
\section{\thesection. Localization properties of Floquet Eigenstates}
\begin{figure}
\begin{center}
\includegraphics[width=1\columnwidth]{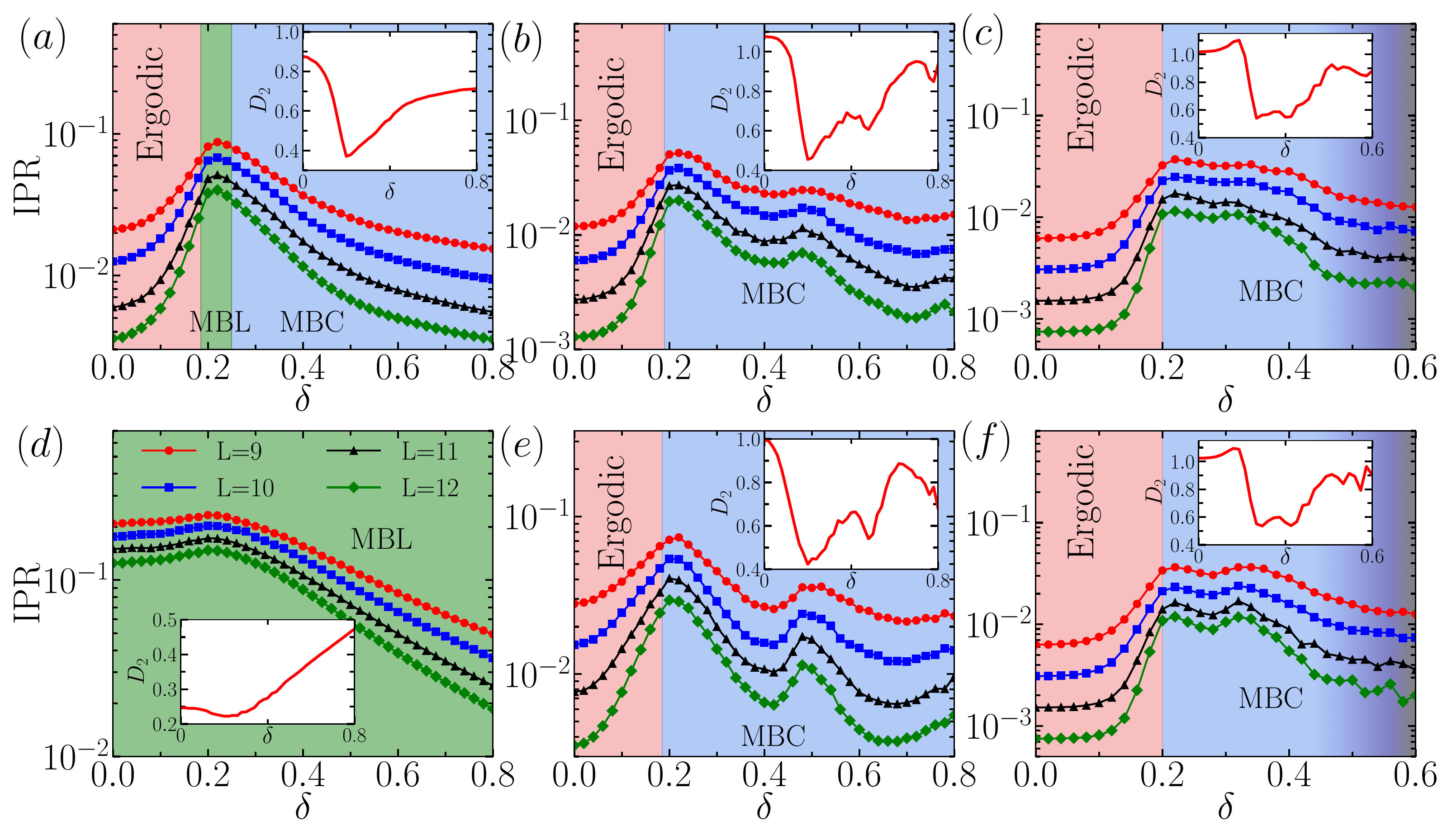}
\end{center}
\caption{(a-f) IPR as a function of $\delta$ for different system sizes $L$. The top (bottom) panel corresponds to $\lambda = 0.3$ ($0.8$). The left, middle, and right column correspond to $T = 0.02\pi$, $1.5\pi$, and $2\pi$, respectively. Insets show the fractal dimension $D_2$ as a function of $\delta$.}
\label{fig:IPR_D2}
\end{figure}

Here we analyze the localization properties of Floquet eigenstates by calculating the inverse participation ratio (IPR), which is a widely used diagnostic for characterizing many-body localization and delocalization, and is defined as $IPR = \sum_i |\psi_i|^4$, where, $\psi_i = \langle i|\psi\rangle$, $|\psi\rangle$ is the normalized many-body eigenstate and $|i\rangle$ is the computational basis. We extend our analysis to investigate the multifractal properties of the system as functions of the driving frequency and the disorder strength. The fractal dimension is extracted from the scaling of IPR with the dimension of the Hilbert space, defined as
\begin{equation}
    \mathrm{IPR} = \sum_i |\psi_i|^4 \sim A\, \mathcal{N}^{-D_2},
\end{equation}
where $A$ is a constant coefficient and $\mathcal{N} = 2^L$ denotes the Hilbert space dimension. The fractal dimension $D_2$ serves as an effective indicator of different phases: $D_2 \simeq 1$ corresponds to a ergodic phase whereas $0<D_2<1$ signals non-ergodic phases including MBL and MBC regimes~\cite{mace2019multifractal,roy2025many}. However, the MBL and MBC can be distinguished from the LSR and also $D_2^{MBC}> D_2^{MBL}$. We extract $D_2$ by fitting the scaling relation
\begin{equation}
    \ln \langle \mathrm{IPR} \rangle = -D_2 \ln(\mathcal{N}) + c,
\end{equation}
using system sizes $L = 9$ to $12$, where $\langle \mathrm{IPR} \rangle$ denotes disorder-averaged IPR. We have plotted the IPR as a function of $\delta$ for different system sizes, and the inset shows the corresponding fractal dimension $D_2$. The Fig.~\ref{fig:IPR_D2}(a) and Fig.~\ref{fig:IPR_D2}(d) shows the IPR as a function of $\delta$ for $T=0.02\pi$, $\lambda=0.3$ and $T=0.02\pi$, $\lambda=0.8$, respectively, corresponding to LSR shown in Fig.~2(a) and Fig.~2(b) in the main text. Combine the IPR and $D_2$ from Fig.~\ref{fig:IPR_D2}(a), we can say that the region $\delta<\mu = 0.2$ is ergodic in nature and $\delta>\mu = 0.2$ is nonergodic phase. By comparing with LSR analysis from Fig.~2(a) of the main text, we conclude that the nonergodic phase consists of MBL phase for $\delta<0.25$ and MBC phase for $\delta>0.25$. However, if we increase the strength of the disorder i.e. $\lambda = 0.8$, which is the case for Fig.~\ref{fig:IPR_D2}(d), we can see the system becomes MBL irrespective of the value of $\delta$ within the defined parameter regime. The minimum of $D_2$ at $\delta=0.2$ is consistent with the location of the Widom line.

Fig.~\ref{fig:IPR_D2}(b) and Fig.~\ref{fig:IPR_D2}(e) show similar plots, but now for $T=1.5\pi$ (low frequency) which are complementary to Figs.~2(c-d) in the main text. Both the figures depict an ergodic-MBC transition at $\delta=\mu=0.2$, denoted by a strong peak (dip) in IPR ($D_2$), followed another weaker peak (dip) in the same at around $(\mu + \delta)T \approx \pi$ tracing the crossover line in the MBC phase.
Plots for even lower frequency $\omega=1$ $(T=2\pi)$ are shown in
Fig.~\ref{fig:IPR_D2}(c) and Fig.~\ref{fig:IPR_D2}(f) for fixed $\lambda=0.3$ and $\lambda=0.8$, respectively.
From both the figures, the ergodic-MBC transition points remains the same whereas the crossover line within the MBC phase shifts towards lower value of $\delta$ according to $(\mu + \delta)T\approx \pi$. We notice that the peak in IPR within MBC phase gets weaker with lowering the driving frequency strength. However, for the fixed frequency, the same becomes more prominent with increasing $\lambda$. We also observe that $D_2$ in the ergodic phase fluctuates around $1$ due to fluctuating data and finite size effects. A proper data-collapse using finite size scaling method leads to $D_2=1$~\cite{roy2025many}.

\setcounter{section}{4}
\section{\thesection. dynamical behavior at lower frequency}
\begin{figure}
\begin{center}
\includegraphics[width=0.8\columnwidth]{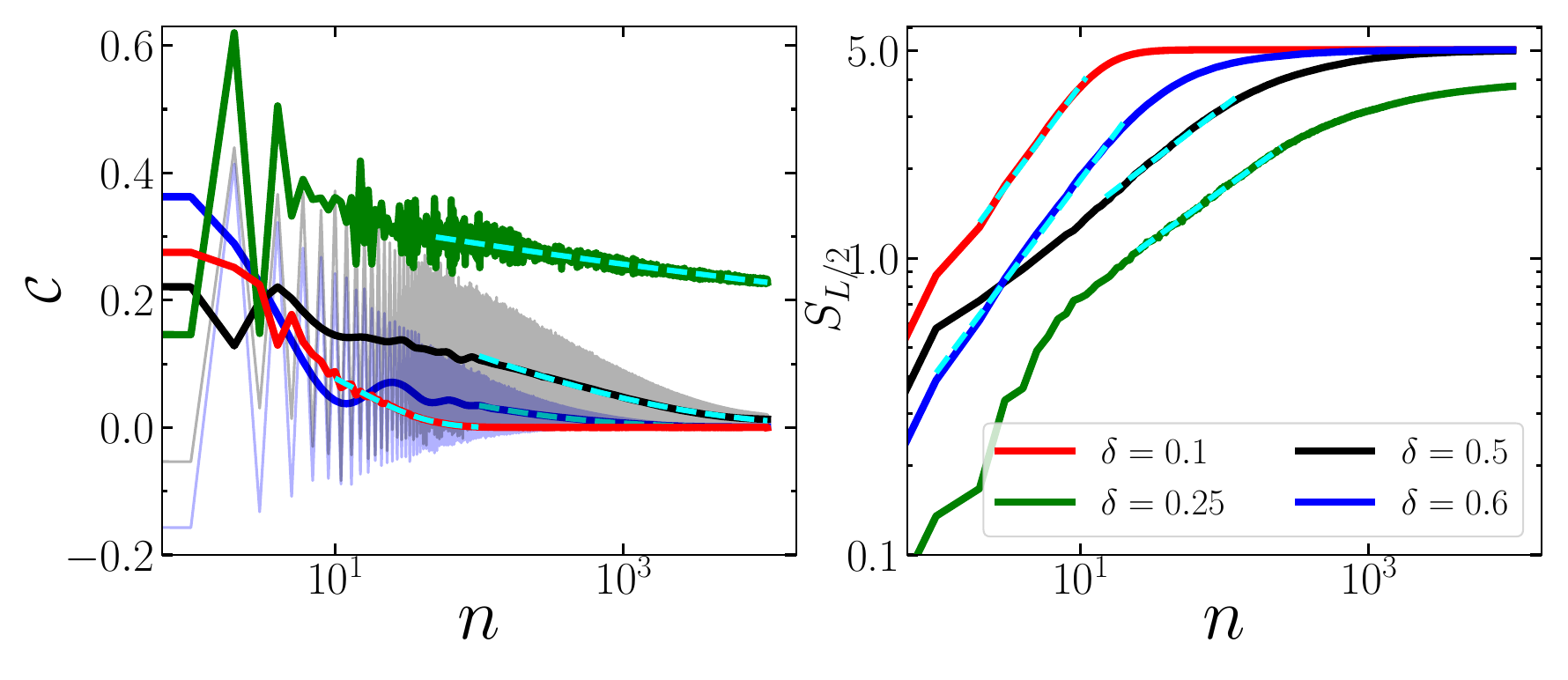}
\end{center}
\caption{Figures (a) and (b) show the autocorrelation $\mathcal{C}$ and half-system EE $(S_{L/2})$, respectively, as a function of the Floquet-step $n$ for driving period $T = 2\pi$. We keep the other parameter fixed at $\mu = 0.2$, $J = 0.2$ and $\lambda = 0.7$. Fitted curves are shown as dashed regions. For all plots $L=16$}.
\label{fig:dynamics_T_2pi}
\end{figure}

Here we provide dynamical characterization of the phase diagram shown in Fig.~\ref{fig:phase_diag_LSR_2Pi}(a) for even larger driving period $T=2\pi$ through the autocorrelation $\mathcal{C}$ and half-chain entanglement entropy $S_{L/2}$ as defined in the main text. For $\delta<0.2$, the system is in the ergodic regime whereas for $\delta>0.2$ it is MBC with the crossover line appearing at $\delta\approx 0.32$. The plots of $\mathcal{C}$ and $S_{L/2}$ are shown in Fig.~\ref{fig:dynamics_T_2pi}(a) and Fig.~\ref{fig:dynamics_T_2pi}(b), respectively, for a fixed $\lambda=0.7$ and different values of $\delta$. At  $\delta=0.1$, $\mathcal{C}$ decays almost exponentially (a stretched exponential fit with $\mathcal{C}\propto \exp[-(n/\tau)^{\gamma_e}]$ reveals $\gamma_e\approx 0.8$) with number of drive $n$ whereas $S_{L/2}$ grows sub-ballistically with power-law exponent $\beta\approx 0.66$, indicating the presence of ergodic phase.
For $\delta=0.25$ in the MBC phase, $\mathcal{C}$ decays in a power-law fashion as $\mathcal{C}\propto n^{-\gamma_p}$ where $\gamma_p \approx 0.05$ and $S_{L/2}$ grows with $\beta\approx0.35$ in sub-diffusive manner. Beyond the crossover line, at $\delta=0.5$, $\mathcal{C}$ decays with $\gamma_e\approx 0.22$ to zero and $S_{L/2}$ slowly grows with $\beta\approx0.37$ to the ergodic value of saturation. At even higher $\delta=0.6$, $\gamma_e \approx0.25$ and $\beta\approx0.65$ with $\mathcal{C}$ and $S_{L/2}$ showing similar longtime behavior as that for $\delta=0.5$. This kind of dynamical behaviors beyond crossover line at large $\delta$ indicates that the regime may be prethermal and may become thermal as well with further lowering the driving frequency.

\setcounter{section}{5}
\section{\thesection. Floquet-Magnus Expansion}
We consider our kicked Ising chain with Floquet operator
\begin{equation}
\hat{U}_F = e^{-i \frac{T}{2} \hat{H}_z} e^{-i \frac{T}{2} \hat{H}_x},
\end{equation}
where
$\hat{H}_x = \sum_i g_i \sigma_i^x$, 
$\hat{H}_z = J \sum_i \sigma_i^z \sigma_{i+1}^z + \sum_i h_i \sigma_i^z.$
The Floquet Hamiltonian $\hat{H}_F$ is defined via
$U_F = e^{-i \hat{H}_F T}.$
Defining
$A = -i\frac{T}{2}\hat{H}_z, B = -i\frac{T}{2}\hat{H}_x,$
we use the Baker-Campbell-Hausdorff expansion
\begin{eqnarray}
\ln(e^A e^B) &= A + B + \frac{1}{2}[A,B] 
+ \frac{1}{12}[A,[A,B]] + \frac{1}{12}[B,[B,A]] + \frac{1}{12}[A,[A,B]] - \frac{1}{24}[B,[A,[A,B]]] + \mathcal{O}(T^5).
\end{eqnarray}
Identifying $\hat{H}_F = \frac{i}{T} \ln(e^A e^B)$, we obtain
\begin{equation}
\hat{H}_F = \hat{H}^{(0)} + \hat{H}^{(1)} + \hat{H}^{(2)} + \hat{H}^{(3)} + \mathcal{O}(T^4).
\end{equation}.
Substituting A and B gives:\\ $$\hat{H}_F
=\frac{\hat{H}_z+\hat{H}_x}{2}
-\frac{iT}{8}[\hat{H}_z,\hat{H}_x]
-\frac{T^2}{96}\Big([\hat{H}_z,[\hat{H}_z,\hat{H}_x]]+[\hat{H}_x,[\hat{H}_x,\hat{H}_z]]\Big)
-\frac{iT^3}{384}[\hat{H}_x,[\hat{H}_z,[\hat{H}_z,\hat{H}_x]]]
+\mathcal O(T^4)$$.

Hence, the leading (zeroth order) term is
\begin{equation}
\hat{H}^{(0)} = \frac{1}{2}(\hat{H}_x + \hat{H}_z) =
\frac{J}{2} \sum_i \sigma_i^z \sigma_{i+1}^z
+ \frac{1}{2} \sum_i g_i \sigma_i^x
+ \frac{1}{2} \sum_i h_i \sigma_i^z.
\end{equation}

The commutator
$[\hat{H}_z,\hat{H}_x] = 2i\sum_j g_j a_j \sigma_j^y$, where the dressed local field 
$a_j \equiv h_j+J(\sigma_{j-1}^z+\sigma_{j+1}^z)$.

\begin{equation}
\hat{H}^{(1)} = -\frac{iT}{8}[\hat{H}_z,\hat{H}_x]=
\frac{T}{4} \sum_i \Big[
J \big(g_i \sigma_i^y \sigma_{i+1}^z + g_{i+1} \sigma_i^z \sigma_{i+1}^y\big)
+ h_i g_i \sigma_i^y
\Big].
\end{equation}

The second-order correction is
\begin{eqnarray}
\hat{H}^{(2)} &&=
-\frac{T^2}{96}
\left(
[\hat{H}_z,[\hat{H}_z,\hat{H}_x]] + [\hat{H}_x,[\hat{H}_x,\hat{H}_z]] \right)
\nonumber \\
&&= -\frac{T^2}{24}\sum_j\Big(g_j a_j^2 \sigma_j^x+g_j^2 a_j \sigma_j^z\Big)
+\frac{JT^2}{12}\sum_j g_j g_{j+1}\sigma_j^y \sigma_{j+1}^y) \nonumber \\
&&=
-\frac{T^2}{24}\sum_j g_j(h_j^2+2J^2)\sigma_j^x
 -\frac{JT^2}{12}\sum_j g_j h_j
(\sigma_{j-1}^z\sigma_j^x+\sigma_j^x\sigma_{j+1}^z) 
-
\frac{J^2T^2}{12}\sum_j g_j \sigma_{j-1}^z\sigma_j^x\sigma_{j+1}^z
-\frac{T^2}{24}\sum_j g_j^2 h_j \sigma_j^z \nonumber
\\
&& \hspace{1cm}-
\frac{JT^2}{24}\sum_j (g_j^2+g_{j+1}^2)\sigma_j^z\sigma_{j+1}^z
+\frac{JT^2}{12}\sum_j g_j g_{j+1}\sigma_j^y\sigma_{j+1}^y.
\end{eqnarray}

The third-order contribution is given by
\begin{eqnarray}
\hat{H}_F^{(3)}&=&
-\frac{iT^3}{384}[\hat{H}_x,[\hat{H}_z,[\hat{H}_z,\hat{H}_x]]].  \nonumber \\ 
&=& -
\frac{JT^3}{24}\sum_j g_j g_{j+1}
\Big(
h_j \sigma_j^x \sigma_{j+1}^y
+
h_{j+1}\sigma_j^y \sigma_{j+1}^x
\Big)
-
\frac{J^2T^3}{24}\sum_j g_j g_{j+1}
\Big(
\sigma_{j-1}^z \sigma_j^x \sigma_{j+1}^y
+
\sigma_j^y \sigma_{j+1}^x \sigma_{j+2}^z
\Big)
\end{eqnarray}

\begin{figure}
\begin{center}
\includegraphics[width=0.5\columnwidth]{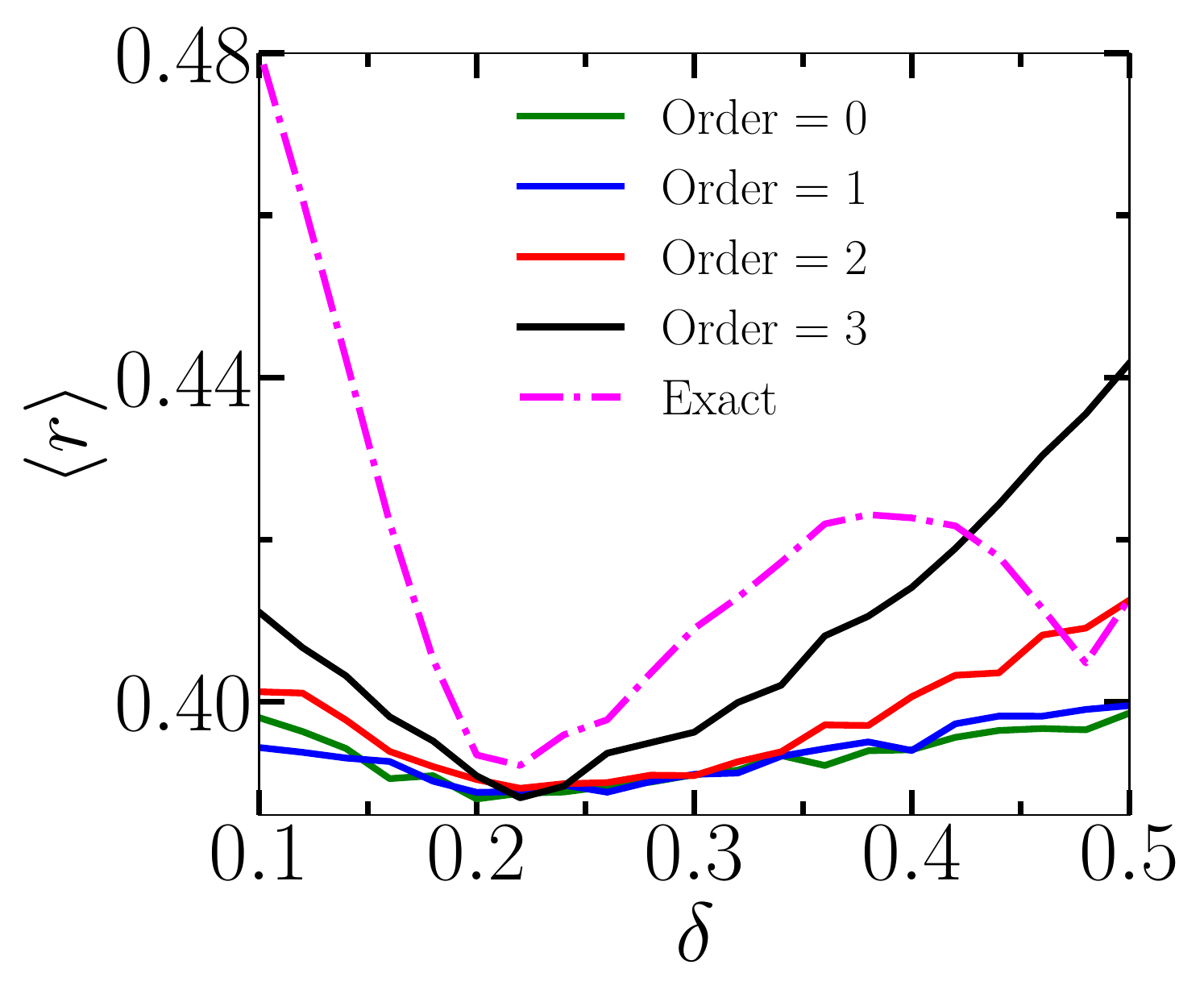}
\end{center}
\caption{Spectrum averaged LSR ($\langle r \rangle$) as a function of $\delta$ for driving period $T = 1.5\pi$, considering the Floquet Hamiltonian~($H_F$) derived from the Floquet-Magnus expansion terminated at increasing orders as shown in the plots . We keep the other parameter fixed at $\mu = 0.2$, $J = 0.2$ and $\lambda = 0.7$ and system size $L=11$, the exact numerical plots for which is shown by the dot-dashed line.}
\label{fig:FM}
\end{figure}

We note that while $H^{(0)}$ retains the structure of high-frequency quasiperiodic Ising chain. For $\lambda=0.7$, the system described by $H^{(0)}$ is in MBL phase. Hence, $\langle r\rangle$ assumes the Poisson value $\approx 0.386$ as shown in Fig.~\ref{fig:FM}. The next higher-order $H^{(1)}$ contains interaction-assisted single spin-flip terms which presumably renormalize the local spin-flip amplitudes but do not contribute much to the Fock-space connections. This is captured as $\langle r\rangle$ do not change much from Poisson value after adding $H^{(1)}$. However, $H^{(2)}$ gives rise to double spin-flip ($\sigma^y\sigma^y$) interaction  terms along with renormalized $\sigma^x$, $\sigma^z$ and $\sigma^z\sigma^z$ terms. This generates more hybridization and hence $\langle r\rangle$ starts increasing from $0.386$ indicating destabilization of the MBL phase at lower frequencies. The next order $H^{(3)}$ also contributes to more hybridization through the double spin-flip ($\sigma^y\sigma^y$) interaction terms and hence $\langle r\rangle$ increases even more in Fig.~\ref{fig:FM}. 
Overall, the Floquet–Magnus expansion reveals that decreasing the driving frequency systematically generates interaction-assisted transverse processes and multi-spin couplings that enhance many-body hybridization. While these processes destabilize the MBL phase by increasing Hilbert-space connectivity, the underlying fragmented structure of Fock-space (due to peculiar structure of these quasiperiodic terms in expansion) prevents complete thermalization, leading instead to a broad nonergodic extended (MBC) regime.

However, the large difference between the exact value of $\langle r\rangle$ and the same obtained from Magnus expansion till 4th order, especially for $\delta<0.2$, tells us that the regime of expansion at $T=1.5\pi$ is highly non-perturbative. Hence, the expansion is not quantitatively controlled at $T=1.5\pi$ and it may not converge for $\delta<0.2$ (ergodic phase) as $\langle r\rangle$ is far way from the exact value. However, for $\delta\geq 0.2$, in the MBC phase, the expansion may converge as $\langle r\rangle$ approaches rapidly toward the exact value although it fails to capture the non-monotonic variation within the same phase.  We use our derivation primarily to identify the operator processes generated by the drive that can nearly justify the destruction of MBL and emergence of ergodic and MBC regimes for $\delta<0.2$ and $\delta>0.2$, respectively, at $T=1.5\pi$.   

\setcounter{section}{6}
\section{\thesection. Analysis of the non-interacting limit}
For $\lambda=0$, the Hamiltonians $\hat{H}_x$ and $\hat{H}_z$ consisting the Floquet drive describe free fermion chains via Jordan-Wigner transformation which we will explain later. For $\lambda>0$, $\hat{H}_z$ describes a chain of interacting fermions. Hence, it is worthwhile to explore the localization properties of eigenstates of our system in the non-interacting limit which give rise to the phase diagram,  shown in the Fig.~1 of the main text, through interaction-induced mixing of the single particle eigenstates in the non-interacting limit.
Here we will discuss the phases on the $\delta$ axis, i.e., when $\lambda=0$. In this limit, the unitary time evolution operator for single Floquet step is given by,
\begin{align}
    \hat{U}_F &= e^{-\frac{iJT}{2}\sum_j\hat{\sigma}_j^z\hat{\sigma}_{j+1}^z} e^
    {-\frac{iT}{2}\sum_jg_j\sigma_j^x}.
    \label{eq:Floquet_uni_lam_0}
\end{align}
Where, each term carries a similar meaning as defined in the main text. It is convenient to perform spin axis rotation before Jordan-Wigner transformation, under such a unitary rotation, $\hat{\sigma}^x \rightarrow \hat{\sigma}^z$ and $\hat{\sigma}^z \rightarrow -\hat{\sigma}^x$. Therefore, the unitary operator given in Eq.~\ref{eq:Floquet_uni_lam_0}, can be written as,
\begin{align}
    \hat{U}_F &= e^{-\frac{iJT}{2}\sum_j\hat{\sigma}_j^x\hat{\sigma}_{j+1}^x} e^
    {-\frac{iT}{2}\sum_jg_j\sigma_j^z}.
\end{align}
This model can be easily solved by transforming the spin operators to fermionic operators using the Jordan-Wigner transformation~\cite{Jordan1928},
\begin{align}
    \sigma_j^- = \prod_{l=0}^{l-1} e^{i\pi \hat{c}_l^\dagger \hat{c}_l}\hat{c}_j, \ \ \ 
    \sigma_j^z = 1 - 2\hat{c}_j^\dagger \hat{c}_j 
\end{align}
Where $\hat{\sigma}_j^x = (\hat{\sigma}_j^+ + \hat{\sigma}_j^-)$. Therefore, the Floquet operator in terms of fermionic operators can be written as,
\begin{align}
    \hat{U}_F &= e^{-\frac{iJT}{2}\sum_j (\hat{c}_j^{\dagger} + \hat{c}_j)(\hat{c}_{j+1}^{\dagger} + \hat{c}_{j+1})} e^
    {-\frac{iT}{2}\sum_jg_j(1-2\hat{c}_j^\dagger \hat{c}_j)}.
    \label{kitaev}
\end{align}
An established way to distinguish delocalized, localized and critical single particle eigenstates in quasiperiodic systems is given by the analysis of the eigen-energies $E_n$ (arranged in ascending order) with PBC. One can define even-odd (odd-even) spacings $s^{e-o}_n=E_{2n}-E_{2n-1}$ ($s^{o-e}_n=E_{2n+1}-E_{2n}$). $s^{e-o}_n$ and $s^{o-e}_n$ are gapped, overlapped and scattered in the delocalized, localized and critical regimes, respectively~\cite{deng2019one}.
\begin{figure}
\begin{center}
\includegraphics[width=1.0\columnwidth]{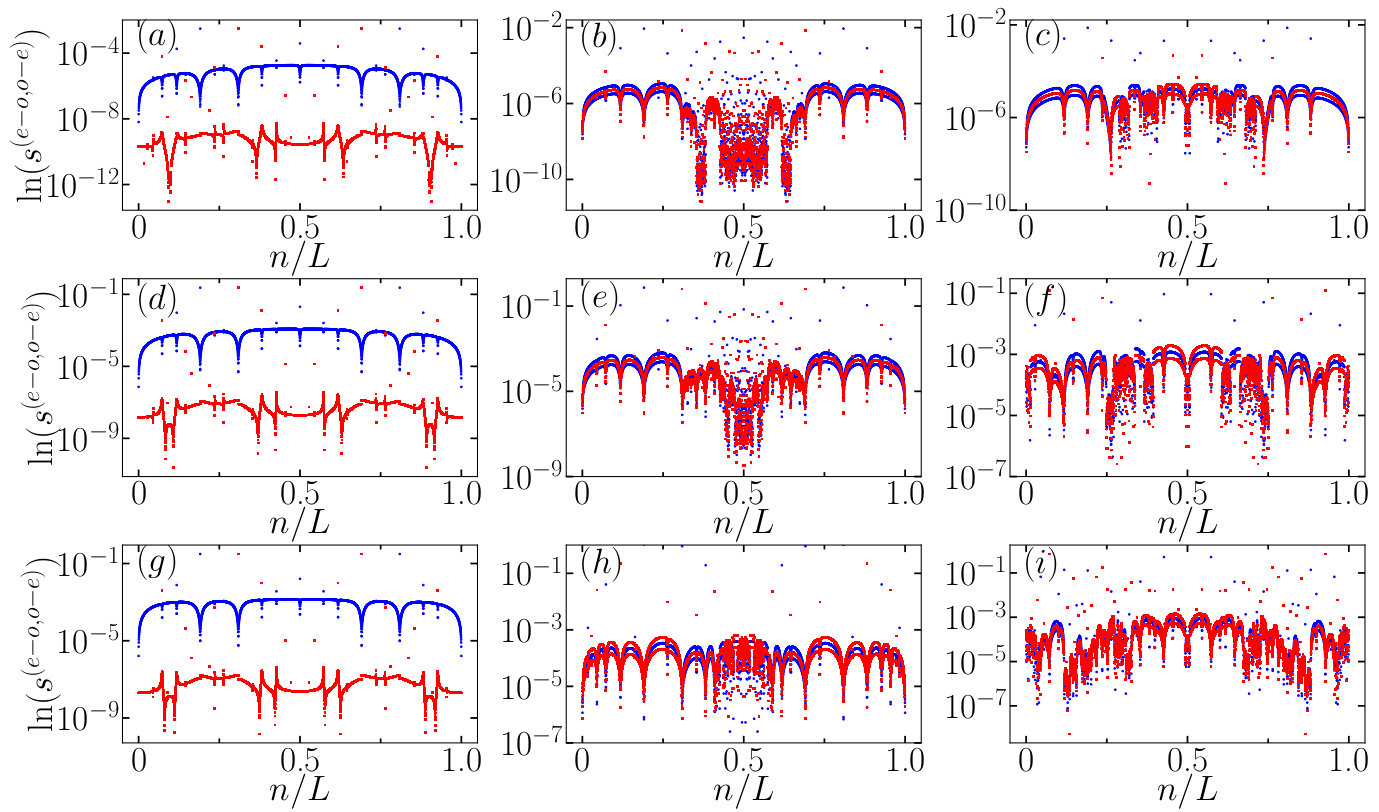}
\end{center}
\caption{ Single particle eigenvalue spacing analysis for $\lambda=0$ model. (a-f) Even-odd and odd-even spacings as a function of $n/L$ with eigenstate index $n$ and system size $L=4181$. The top, middle and bottom panels correspond to $T=0.02\pi, 1.5\pi$ and $2\pi$, respectively. $\delta=0.1,0.3$ and $0.6$ for the left, middle and right columns, respectively.}
\label{fig:even_odd_spec}
\end{figure}

We employ the same analysis, as described above, for quasi-energies of the driven system in Eq.~\ref{kitaev}. Our results are shown in Fig.~\ref{fig:even_odd_spec} where top, middle and bottom panels correspond to time periods $T=0.02\pi,1.5\pi$ and $2\pi$, respectively.
We find that for $\delta<0.2$ all states are delocalized for all frequencies which are depicted in Figs.~\ref{fig:even_odd_spec}(a), (d) and (g). However, for $\delta>0.2$ we observe presence of localized and critical states separated by anomalous mobility edges, the location of which varies as a function of $T$ and $\delta$. As can be seen from Figs.~\ref{fig:even_odd_spec}(b), (e) and (h), there are critical states at the central part of the spectrum whereas Figs.~\ref{fig:even_odd_spec}(c), (f) and (i) show the critical states which are more scattered in the spectrum. We also notice for $T=0.02\pi$ and $T=1.5\pi$ the spectra do not change much. However, at $T=2\pi$, we observe that for $\delta=0.3$ the fraction of critical states has decreased whereas for $\delta=0.6$ the same fraction has increased as compared to the lower $T$ counterparts. 

In presence of non-zero finite $\lambda$, the effective interaction-induced mixing of the single particle states take place giving rise to the many-body phase diagram discussed in the main text. It is also noteworthy that for $\lambda=0$, the our quasiperiodic spin chain remains integrable, similar to the clean kicked Ising chain~\cite{Prosen_intigrability}, under fast and slow driving. However, due to quasi-periodicity, the many-body eigenstates are ergodic $(D_2=1)$ and critical $(D_2<1)$ for $\delta<0.2$ and $\delta\geq 0.2$, respectively, with no signature of enhanced localization within the critical phase, even in the low-frequency regime. Interestingly, for $\lambda>0$ the integrability gets broken and one obtains the enhanced localization within MBC phase tracing the crossover line.

\end{document}